\documentclass[twocolumn,showpacs,amsmath,amssymb,superscriptaddress,prl]{revtex4}

\usepackage{graphicx}
\usepackage{dcolumn}
\usepackage{bm}
\usepackage{indentfirst}

\begin{document}


\title{Electron transport in single wall carbon nanotube weak links in the Fabry-Perot regime}

\author{H. I. J\o rgensen$\dagger$}
\email{hij@fys.ku.dk}
\affiliation{Nano-Science Center, Niels Bohr
Institute, University of Copenhagen, Universitetsparken 5,
DK-2100~Copenhagen \O , Denmark}
\author{K. Grove-Rasmussen$\dagger$}
\affiliation{Nano-Science Center, Niels Bohr Institute, University
of Copenhagen, Universitetsparken 5, DK-2100~Copenhagen \O ,
Denmark}
\author{T. Novotn\' y}
\affiliation{Nano-Science Center, Niels Bohr Institute, University
of Copenhagen, Universitetsparken 5, DK-2100~Copenhagen \O ,
Denmark}
\affiliation{Department of Electronic Structures, Faculty
of Mathematics and Physics, Charles University, Ke Karlovu 5, 121 16 Prague, Czech Republic\\
$\dagger$ These authors contributed equally to this work}
\author{K. Flensberg}
\affiliation{Nano-Science Center, Niels Bohr Institute, University
of Copenhagen, Universitetsparken 5, DK-2100~Copenhagen \O ,
Denmark}
\author{P. E. Lindelof}
\affiliation{Nano-Science Center, Niels Bohr Institute, University
of Copenhagen, Universitetsparken 5, DK-2100~Copenhagen \O ,
Denmark}

\date{\today}

\begin{abstract}
We fabricated reproducible high transparency superconducting
contacts consisting of superconducting Ti/Al/Ti trilayers to gated
single-walled carbon nanotubes (SWCNTs). The reported semiconducting
SWCNT have normal state differential conductance up to $3e^2/h$ and
exhibit clear Fabry-Perot interference patterns in the bias
spectroscopy plot. We observed subharmonic gap structure in the
differential conductance and a distinct peak in the conductance at
zero bias which is interpreted as a manifestation of a supercurrent.
The gate dependence of this supercurrent as well as the excess
current are examined and compared to a coherent theory of
superconducting point contacts with good agreement.
\end{abstract}

\pacs{74.45.+c, 73.23.Ad, 73.63.Fg, 74.50.+r}

\keywords{MAR AND nanotube AND excess current AND Fabry-Perot AND
Josephson junction}

\maketitle

Electron transport through a SWCNT bridging two metal electrodes has
been studied intensively over the last years. At low temperatures
different transport regimes depending on the transparency of the
metal-SWCNT interfaces have been identified. With low transparency
contacts a quantum dot (QD) will be defined in the SWCNT
\cite{Tans,Bockrath,NygaardQD,Zant}, and with intermediate
transparency contacts Kondo resonances around zero bias are observed
\cite{Nygaard,schonenberger}. High transparency contacts are in
recent years also reported, where the SWCNT constitutes an electron
waveguide with Fabry-Perot (FP) interferences \cite{Liang,Dai}.

Changing the metal electrodes to superconductors (SC) dramatically
changes the transport characteristics. In such junctions the carbon
nanotube (CNT) forms a weak link between the two superconductors and
several interesting effects can be observed. At zero bias a
supercurrent can flow through the weak link \cite{pablo, haruyama,
Kasumov} due to the Josephson effect \cite{tinkhambog}, while at low
biases current will be carried by Multiple Andreev reflections (MAR)
at the two CNT-S interfaces
\cite{morpurgo02,buitelaar,buitelaarkondo}. For large bias, these
effects will give rise to an excess current.

In fact, only very recently have these effects been seen in SWCNT
devices \cite{pablo} similar to the one presented here. In this
Letter, we present transport measurements on a gated S-SWCNT-S
Josephson junction at low temperatures, with high transparency
contacts. We focus on the gate dependence of the excess current and
zero-bias conductance peak in the FP regime.

The SWCNTs are grown by chemical vapor deposition (CVD) from
catalyst islands made by electron beam lithography and positioned
relative to predefined alignment marks. The details of the
CVD-growth procedure are described elsewhere \cite{Kong, kgrhij}.
After growth, source and drain electrodes consisting of
superconducting trilayers are positioned next to the catalyst
islands to contact the SWCNT. The gap between the source and drain
trilayer films is approximately 500 nm. The superconducting
trilayers consist of 5 nm titanium to make good contact to the
SWCNT, then of 40 nm aluminum to raise the transition temperature,
and finally 5 nm titanium to stop oxidation of the aluminum.

Our devices are made on a highly doped silicon wafer with a
0.5\,$\mu$m thermally oxidized SiO$_{2}$ layer on top. We use the
silicon substrate as a back gate. To be able to measure the
transition temperature $T_{C}$ and the critical field $B_{C}$ of the
trilayer films at low temperatures we furthermore define a four
probe device of the superconducting trilayer. For the device in this
Letter we find $T_{C}=750$\,mK, $B_{C}=75$\,mT and from BCS theory
we calculate a superconducting energy gap of $2\Delta=3.5 k_{B}
T_{C}=230$\,$\mu$eV. However, the actual effective value of $\Delta$
for the CNT weak link might differ from this measured value due to
interface effects and, indeed, we found from the fit of the excess
current measurements (see below) that the effective gap is reduced
by about a factor of $\sim 0.7$. All measurements are preformed at
$300\,{\rm mK}$ in a sorption pumped ${}^3$He cryostat (Oxford
Instruments Heliox). The measurements are made with standard DAQ
cards, lock-in amplifiers (excitation 5\,$\mu$V), and opto-couplers
to reduce noise.

\begin{figure}
\begin{center}
\includegraphics[width=0.48\textwidth]{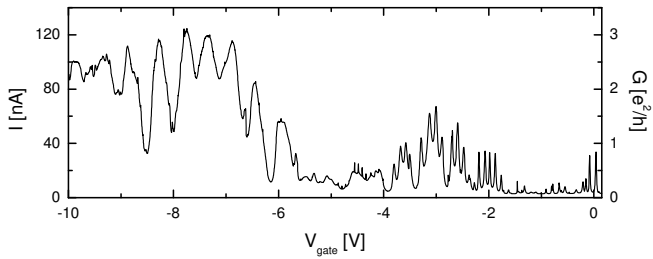}
\end{center}
\caption{Current and conductance for the S-SWCNT-S Josephson
junction as a function of voltage applied to the back gate with
$1\,$mV source drain voltage. The high conductance regime with
Fabry-Perot oscillations is reached at large negative gate
voltages ($<-6\,$V). Coulomb blockade peaks are seen at lower gate
voltages.}
\label{fig:gatesweep}
\end{figure}

Fig.~\ref{fig:gatesweep} shows a gatesweep from -10\,V to 0\,V with
$V_{sd}=1\,{\rm mV} ~(>2\Delta/e)$. It displays strong gate
dependence: High conductance at high negative gate voltages and low
conductance at small gate voltages, which indicates that the SWCNT
is semiconducting. Addition of the first hole and first electron
cannot be identified, probably due to substrate perturbations. The
SWCNT defines a QD with gate depended Schottky barriers at each
interface. The gate thus tunes both the energy levels of the QD and
the strength of the Schottky barriers.

From $V_{\rm gate}\sim\,-4\,$V to $V_{\rm gate}\sim-2\,$V the
Schottky barriers are large and the SWCNT constitutes a closed QD,
{\em i.e.}, the charging energy $U_{C}$ is larger than the
broadening of the energy levels $\Gamma$. Transport is here
dominated by charging effects and Coulomb blockade peaks are clearly
visible. Some of the peaks are spaced into periods of four due the
four-fold degeneracy (spin and orbital) of each energy level, also
confirmed by bias spectroscopy plots (not shown). Such
characteristic is sign of a high quality SWCNT. As the gate voltage
is decreased to more negative values the Schottky barriers are
decreased, leading to an increase of $\Gamma$. Below $V_{\rm
gate}\sim -5\,$V the dot opens, $\Gamma$ becomes larger than
$U_{C}$, and charging effects of the QD are no longer dominant.
Instead FP interference of the electron waves being reflected at the
SWCNT-electrode interfaces dominates transport.
\begin{figure}
\begin{center}
\includegraphics[width=0.48\textwidth]{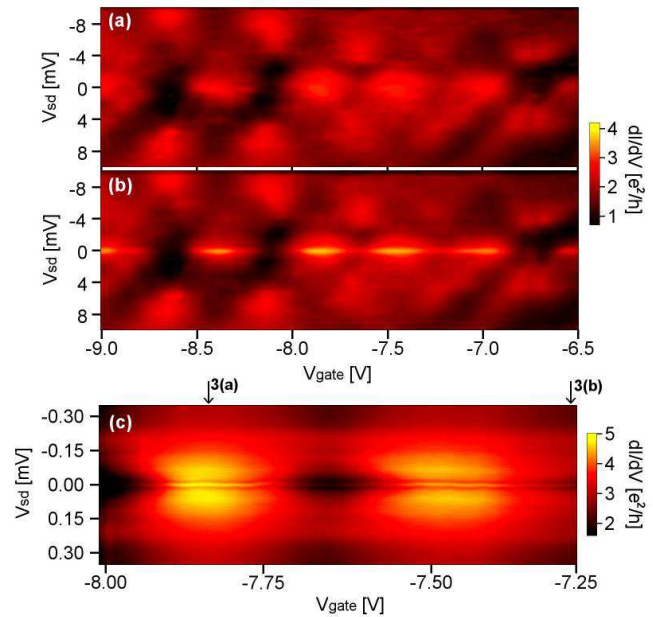}
\end{center}
\caption{(Color online) (a) Bias spectroscopy plot in the high
transparency gate region, with a small magnetic field ($B=100\,$mT)
to suppress the superconducting state of the electrodes. (b)
Analogous to (a) but without magnetic field, {\em i.e.}, with
superconducting electrodes. (c) Close-up on two resonances from
which the excess current and supercurrent of Fig.~\ref{fig:excess}
are extracted. Arrows are pointing to the gate voltages where the
graphs in Figs.~\ref{fig:cut}(a-b) are measured.}
\label{fig:biasspec}
\end{figure}

Fig.~\ref{fig:biasspec}(a) shows a bias spectroscopy plot in this
gate region with a small magnetic field applied ($B=100\,$mT) to
suppress the superconducting state of the electrodes.\\
The average differential conductance in this gate region is around
$\sim 2.5e^2/h$ with maximums of about $\sim 3 e^2/h$, approaching
the theoretical limit of $4e^2/h=(6.5\,{\rm k}\Omega)^{-1}$. The
maximum value of $\sim 3 e^2/h$ of the conductance implies a
rather large asymmetry $\Gamma_L/\Gamma_R\approx 5$ of the CNT
couplings to the contacts (found from a resonant level model
representing the system well around a conductance peak), in
contrast to fairly symmetric couplings reported elsewhere
\cite{buitelaar,pablo}. As $V_{\rm gate}$ and $V_{\rm sd}$ are
changed the dips in conductance evolve into straight lines,
forming a mesh of crossing dark lines. These pronounced
oscillations in differential conductance versus $V_{\rm gate}$ and
$V_{\rm sd}$ are clear signs of the FP interferences \cite{Liang}.

\begin{figure}
\begin{center}
\includegraphics[width=0.48\textwidth]{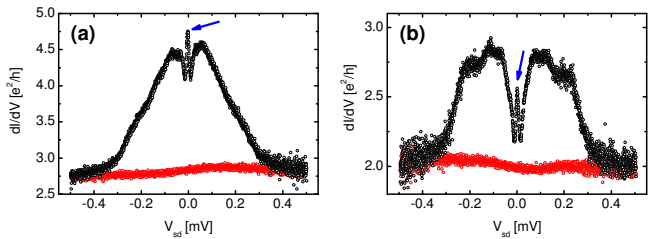}
\end{center}
\caption{(Color online) Differential conductance versus source drain
voltage measured with lock-in amplifier ($5\,\mu$V excitation) at
different gate voltages as indicated in Fig.~\ref{fig:biasspec}(c).
Upper curve (black) is for the superconducting electrodes and lower
curve (red) is with the magnetic field of $100\,$mT to suppress the
superconductivity of the electrodes. (a) is measured at a resonance
of the Fabry-Perot pattern while (b) is measured at a antiresonance.
Blue arrows point to the supercurrent peak. } \label{fig:cut}
\end{figure}

As we turn off the magnetic field, {\em i.e.}, turn on the
superconducting state of the electrodes, an overall increase in
differential conductance between $V_{\rm sd}\sim \pm 2\Delta/e$ is
observed (Fig.~\ref{fig:biasspec}(b)). A more detailed bias
spectroscopy plot of this overall increase through two successive
resonances is shown in Fig.~\ref{fig:biasspec}(c). Detailed
measurements with lock-in amplifier of differential conductance
versus $V_{\rm sd}$ at gate voltages indicated in
Fig.~\ref{fig:biasspec}(c) are shown in Figs.~\ref{fig:cut}(a-b),
where (a) is a cut through a resonance and (b) is a cut through an
antiresonance. A characteristic conductance variation between
$V_{\rm sd}\sim \pm 2\Delta /e$ is seen for all gate voltages.
Close to $|V_{\rm sd}| \sim 2\Delta/e$ the conductance starts to
increase while at smaller source drain voltages a dip centered
around zero bias also develops. In Fig.~\ref{fig:cut}(b) this dip
can be seen between $V_{\rm sd} \sim \pm 80\,\mu$V and less
strongly in Fig.~\ref{fig:cut}(a). The change in conductance
between $V_{\rm sd} \sim \pm 2\Delta/e$ to typically higher, but
sometimes also lower values, than the normal state differential
conductance $G_{N}$ is because superconductivity induced transport
mechanisms occur.

Between $V_{\rm sd}\sim \pm 2\Delta/e$ transport is governed by
Andreev reflections (ARs) \cite{Andreev} and normal reflections. An
electron with energy $|\epsilon| < \Delta$ relative to the Fermi
energy in the normal region has (depending on the barrier strength)
a probability for being AR on the superconductor as a hole
effectively transferring two electrons (one Cooper pair) through the
NS interface \cite{btk,ktb}. For $|\epsilon | > \Delta$ ARs are
still possible but fall off rapidly. Multiple Andreev reflections
between the two superconducting leads at finite bias give rise to a
sub gap structure (SGS) \cite{lindelof,obtk,cuevas,martinrodero},
while at zero bias a dissipationless supercurrent can flow providing
that the interfaces are sufficiently transparent. In
Fig.~\ref{fig:cut}(b) we observe features for $|V_{\rm sd}|\lesssim
2\Delta /e$ and a distinct peak around zero bias, which is a general
trend in our S-SWCNT-S junctions with high transparency. At $|V_{\rm
sd}| \gg 2\Delta/e$ transport is mostly due to quasiparticle
transport and the FP pattern is seen in Fig.~\ref{fig:biasspec}(b).
As we approach $|V_{\rm sd}|\sim 2\Delta/e$ from above the
quasiparticle transport is enhanced due to the modified density of
states of the superconductors and below this point a subharmonic gap
structure (SGS) is expected to appear. We observe a complex pattern
in the subgap region for the lower transparency case
(Fig.~\ref{fig:cut}(b)) while the higher transparency one
(Fig.~\ref{fig:cut}(a)) doesn't show much structure in qualitative
agreement with theoretical predictions \cite{yeyati,martinrodero}.
The structure in Fig.~\ref{fig:cut}(b) seems too smeared to allow
for quantitative comparison with theory, yet it is an interesting
subject for further studies.

\begin{figure}
\begin{center}
\includegraphics[width=0.48\textwidth]{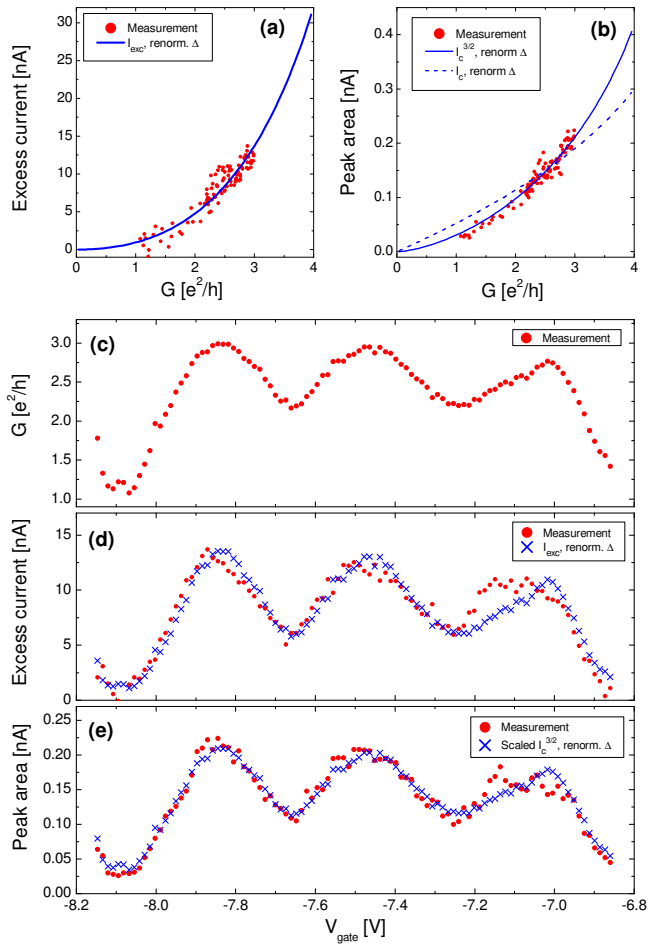}
\end{center}
\caption{(Color online) Comparison between measured and theoretical
data. (a) Measured (dots) and fitted (line) excess current as a
function of conductance (G$ = g\,e^2/h$). The fitted line is
calculated from Eq.~(1) using $\tilde\Delta$ as a fitting parameter.
(b) Measured (dots) and fitted (lines) supercurrent. The fits are to
$a(I_c)^{3/2}$ (full line) and $aI_c$ (dashed line), using $a$ as a
fitting parameter, see text for discussion. (c) Measured conductance
in the Fabry-Perot regime as a function of gate voltage. (d)
Measured (dots) and calculated (crosses) excess current. (e) The
zero bias anomaly area measured (dots) and calculated (crosses). The
theoretical points (crosses) in (d) and (e) are calculated using the
measured conductance in (c) and Eqs.~(1) and (2), respectively. }
\label{fig:excess}
\end{figure}

Instead of studying in detail the SGS, we focused on its integral
effect in the form of the excess current, $I_{\rm exc}$, which is
defined as the difference in current between having the electrodes
in the superconducting state and the normal state at $V_{\rm
sd}\gg\Delta/e$. It can therefore be found as half of the difference
in area between the two set of data points in
Figs.~\ref{fig:cut}(a-b). In Fig.~\ref{fig:excess}(a), we have
extracted the excess currents from the bias spectroscopy plot in
Fig.~\ref{fig:biasspec}(c) and plotted them as functions of the
normal state differential conductance. Since the level width is much
larger than the SC gap, $\Gamma\gg\Delta$, as can be seen from
Figs.~\ref{fig:biasspec},~\ref{fig:cut} we can use for the
interpretation of the results the well-established theory of
superconducting (SC) point contacts \cite{shumeiko,yeyati} and fit
the excess current with the function
\begin{equation}
I_{\rm exc}(g)=\!\frac{e\tilde{\Delta}\,g^2}{h(4-g)}\bigg[1 -
\frac{g^2}{4\sqrt{4 - g}(8 - g)} \log\frac{2 + \sqrt{4 - g}}{2 -
\sqrt{4 - g}}\bigg]
\end{equation}
where $g$ is the conductance measured in units of $e^2/h$ and where
$\tilde{\Delta}$ is the gap parameter at the superconductor-CNT
interface. Allowing for renormalization of $\tilde{\Delta}$ and
performing a least-square fit to the measured data, we get
$\tilde{\Delta}\sim 0.7\Delta$. Using this value in Eq.~(1) yields
the curve in Fig.~\ref{fig:excess}(a) and also the data points in
Fig.~\ref{fig:excess}(d), showing good agreement between experiments
and the theoretically extracted excess current.

Next, we discuss the zero bias anomaly that we observe for all gate
voltages in the FP-region. The peak has a full width of only $\sim
25\,\mu$V (see Fig. \ref{fig:cut}). Such a peak in conductance at
zero bias was observed earlier \cite{buitelaar,buitelaarkondo} and
was attributed to a dissipative quasiparticle current
\cite{vecinoinelastic}. In this work, however, we pursue an
alternative interpretation viewing the peak as a manifestation of
the supercurrent. We interpret half of the area of the zero-bias
peak ({\em not} the whole area {\em under} the peak) as a measure of
the supercurrent. At first glace, this interpretation seem
inconsistent since the measured peak area is of the order of $0.2$
nA, while the expected magnitude of the supercurrent is on the order
of $2e\tilde{\Delta}/\hbar\sim 35$ nA, {\em i.e.}, more than two
orders of magnitude more. However, a similar discrepancy between the
measured and expected values of the supercurrent has been observed
previously \cite{pablo,devoret}. In particular, in a very recent
study \cite{pablo} on a similar device to ours where the
supercurrent was measured as zero bias current a discrepancy of the
order of $15$ between the measured and expected value was found. The
suppression can be understood in terms of the dynamics of the SC
phase of a resistively shunted Josephson junction with the
environment, resulting in an apparent critical current $I_{m}$
scaling as $I_{m}\propto I_{c}^{3/2}$ with $I_c$ being the bare
critical value of the supercurrent $I_c$, see Refs.~\cite{pablo,
devoret} for details \footnote{The important parameter for the
validity of this description is the quality factor $Q > 1$ which is
estimated to be 2 in our case \cite{pablo}.}.

Adopting this idea, we perform a fit to the measured zero-bias-peak
area as a function of the normal state conductance,
Fig.~\ref{fig:excess}(b). The supercurrent is determined as
\cite{martinrodero,shumeiko}
\begin{equation}\label{Ic}
I_c(g)\!=\!\frac{e\tilde{\Delta}\,
g\sin\varphi_\mathrm{max}}{4\hbar\sqrt{1-\frac{g}{4}\sin^2(\frac{\varphi_\mathrm{max}}{2})}}
\tanh\frac{\tilde{\Delta}\sqrt{1-\frac{g}{4}\sin^2(\frac{\varphi_\mathrm{max}}{2})}}{2k_BT}
\end{equation}
where $\varphi_\mathrm{max}$ is the phase at which the supercurrent
in Eq.~\eqref{Ic} is maximal. Using the renormalized value of the
gap, we have fitted the data to both $a I_c(g)$ and $a I_c^{3/2}(g)$
with $a$ being a fitting parameter. We clearly see that the
conductance dependence of $I_c^{3/2}(g)$ fits the measured data very
well. On the other hand the dependence of $I_c(g)$ doesn't fit the
data at all, comparable with analogous results of Ref.~\cite{pablo}.
The resulting $V_{gate}$-dependence of the peak area is plotted in
Fig.~\ref{fig:excess}(e) using the fitted values of $I_c^{3/2}(g)$
with $\tilde{\Delta}$. Thus, we conclude that the zero-bias-peak
results are fully consistent with the theoretical predictions based
on the supercurrent interpretation.

In conclusion, we have successfully fabricated gated S-SWCNT-S
Josephson junctions with high transparency contacts. In the
Fabry-Perot regime of the semiconducting SWCNT reported here we
observed quasiparticle tunneling at $|V_{\rm sd}|=2\Delta/e$,
enhanced current due to MARs for $|V_{\rm sd}|< 2\Delta/e$, and a
conductance peak around zero bias. We interpret the zero bias
conduction peak as a not fully developed supercurrent. The excess
current, which has not been analyzed before for such junctions, fits
very well to the theory of coherent SNS junctions.

We wish to acknowledge the support of the Danish Technical
Research Council (The Nanomagnetism framework program), EU-STREP
Ultra-1D program and the Nano-Science Center, University of
Copenhagen, Denmark. The work of T.~N.\ is a part of the research
plan MSM 0021620834 that is financed by the Ministry of Education
of the Czech Republic.

\bibliography{references}
\end{document}